\documentclass[a4paper,11pt]{article}
\pdfoutput=1

\usepackage{jheppub}
\usepackage{mathtools}
\usepackage{tensor}
\usepackage{color}
\allowdisplaybreaks

\usepackage{tabularx}

\newcommand{\flavor}[1]{\left\langle{#1}\right\rangle}
\newcommand{\commute}[2]{\left[{#1, #2}\right]}
\newcommand{\anticommute}[2]{\left\{{#1, #2}\right\}}

\title{Renormalization of chiral perturbation theory with spinless matter field in curved spacetime}

\author[a,b]{Cheng-Cheng Li,}
\author[a]{Xiong-Hui Cao,} 
\author[a,b]{Feng-Kun Guo}

\affiliation[a]{
    Institute of Theoretical Physics, Chinese Academy of Sciences, Beijing 100190, China}

\affiliation[b]{
    School of Physical Sciences, University of Chinese Academy of Sciences, Beijing 100049, China}

\emailAdd{lichengcheng@itp.ac.cn}
\emailAdd{xhcao@itp.ac.cn}
\emailAdd{fkguo@itp.ac.cn}

\abstract{
We generalize chiral perturbation theory with spinless matter fields in the fundamental representation of ${\rm SU}(N)$ to curved spacetime in the presence of an external gravitational field.
This work is motivated by recent interest in investigating energy-momentum tensor matrix elements of matter fields.
The complete chiral Lagrangian is constructed, including both minimal extensions from flat spacetime and new curvature-induced terms, up to the next-to-next-to-leading order in the chiral expansion, $\mathcal{O}(p^3)$.
We perform a systematic one-loop renormalization of the generating functional using the background field method combined with the heat-kernel technique, and explicitly compute the resulting ultraviolet divergences.}

\begin{document}
\maketitle
\flushbottom

\section{Introduction}
In recent years, gravitational form factors (GFFs), defined as hadronic matrix elements of the energy-momentum tensor (EMT), have attracted significant attention in the particle and nuclear physics communities.
Analogous to electromagnetic form factors, which reveal the spatial distributions of charge and magnetization, GFFs encode the fundamental mechanical properties of hadrons, characterizing the internal distributions of mass, spin, pressure and shear forces; see Refs.~\cite{Polyakov:2018zvc,Lorce:2025oot} for reviews.

The EMT in quantum chromodynamics (QCD), as well as in its low-energy effective field theories, can be rigorously defined by coupling the matter fields to an external gravitational source, $g_{\mu\nu}$, in a covariant manner. This is implemented by promoting the flat-spacetime action to a generally covariant one~\cite{Belitsky:2005qn},
\begin{equation} \label{Eq: generalization from flat to curved spacetime}
S_\text{flat}[\Psi] \to S_\text{curv}[\Psi, g_{\mu\nu}] \: ,
\end{equation}
where $\Psi$ collectively denotes the matter fields and $g_{\mu\nu}$ is the metric tensor in curved spacetime.\footnote{We employ the metric signature $(+,-,-,-)$.}
Following this generalization, the symmetric Belinfant{\'e} EMT is derived as the functional derivative with respect to the metric,
\begin{equation}
    T_{\mu\nu}
=   \frac{2}{\sqrt{-g}}\frac{\delta{S_\text{curv}}}{\delta{g^{\mu\nu}}}\bigg |_{g_{\mu\nu}=\eta_{\mu\nu}} \: ,
\end{equation}
where $g$ denotes the determinant of the metric tensor, and $\eta_{\mu\nu}$ is the Minkowski metric tensor in flat spacetime.
The generalization in Eq.~\eqref{Eq: generalization from flat to curved spacetime} is conventionally implemented in two steps.
First, the minimally coupled action $S_\text{curv}^{(M)}$ is constructed by replacing the Minkowski metric $\eta_{\mu\nu}$ with a general metric $g_{\mu\nu}$ and ordinary derivatives with covariant derivatives in the flat-spacetime action $S_\text{flat}$, thereby ensuring invariance under general coordinate transformations.
Subsequently, non-minimal couplings are introduced to form the full action $S_\text{curv}^{(R)}$, which involve explicit couplings between matter fields and the curvature tensor, its contractions (the Ricci tensor $R_{\mu\nu}$ and the Ricci scalar $R$), and possibly higher-order curvature invariants.
As an illustrative example, consider a free spin-0 boson $\phi$ described by the Klein--Gordon Lagrangian~\cite{Birrell:1982ix},
\begin{align}
     S_\text{curv}=\int\mathrm{d}^4x~\sqrt{-g}\left(\frac{1}{2}g^{\mu\nu}(x)\left(\partial_\mu \phi\right)\left(\partial_\nu \phi\right)-\frac{1}{2} m^2 \phi^2-\frac{1}{2}hR(x)\phi^2\right) ,
 \end{align}
where $-h R(x) \phi^2/2$ is a non-minimal coupling term with $h$ being a dimensionless constant.
Such non-minimal terms are required for consistent renormalization and introduce additional low-energy constants (LECs) in chiral perturbation theory (ChPT).
Crucially, they contribute to the resulting EMT, even though they vanish in the flat-spacetime limit.

The extension of ChPT to curved spacetime provides a powerful, model-independent framework for investigating hadronic EMT matrix elements at low momentum transfers.
The chiral Lagrangians for the Goldstone bosons in curved spacetime were first formulated in Ref.~\cite{Donoghue:1991qv}, which enabled the determination of GFFs for pions, kaons, and $\eta$ mesons; see Ref.~\cite{Kubis:1999db} for their electromagnetic corrections.
For the nucleon, the GFFs were first studied at the tree level in Ref.~\cite{Belitsky:2002jp}.
References~\cite{Diehl:2006ya, Alharazin:2020yjv} provided calculations up to chiral $\mathcal{O}(p^3)$ in the heavy-baryon and covariant extended-on-mass-shell (EOMS) schemes, respectively, where $p$ denotes the small momentum scale in the chiral expansion.
More recently, EOMS-based calculations have been extended to $\mathcal{O}(p^4)$ and include the $\Delta$ resonance as explicit degrees of freedom in the small-scale expansion scheme~\cite{Alharazin:2023uhr}.
Recently, this formalism has been extended to studies of, for example, the $\rho$ meson GFFs~\cite{Epelbaum:2021ahi}, the $\Delta$ resonance GFFs~\cite{Alharazin:2022wjj}, the $p \to \Delta^+$ transition GFFs~\cite{Alharazin:2023zzc}, and the deuteron GFFs~\cite{Panteleeva:2024abz}.
In particular, by matching the precise model-independent nucleon GFFs from a recent dispersive analysis~\cite{Cao:2024zlf} to those obtained from ChPT with an external gravitational source at the next-to-next-to-leading order, Ref.~\cite{Cao:2025dkv} determined two relevant LECs $c_8,c_9$ of the meson--nucleon Lagrangian.

These developments motivate us to consider the renormalization of ChPT with matter fields in curved spacetime using the path integral formulation. As a first step, we focus on ChPT with spinless matter fields in the fundamental representation of ${\rm SU}(N)$ up to $\mathcal{O}(p^3)$.
Such a framework is not only of formal interest, but also has practical applications.
First, it enables the study of kaon graviproduction by treating the kaon as a heavy spinless matter field rather than a Goldstone boson. SU(2) ChPT with the kaon treated as a matter field was first developed in Ref.~\cite{Roessl:1999iu} up to the fourth chiral order. 
Such an approach is expected to exhibit better convergence of the chiral expansion compared to ${\rm SU}(3)$ ChPT, given the relatively large mass of the strange quark, albeit at the price of introducing additional LECs. 
Kaon graviproduction, which produces kaons via the energy-momentum tensor, parallels pion graviproduction but provides complementary insights into ${\rm SU}(3)$ flavor symmetry breaking~\cite{Xu:2023izo,Cao:2025dkv,Liu:2025vfe}, and may offer constraints on new LECs through hard exclusive processes~\cite{Aguilar:2019teb, AbdulKhalek:2021gbh}.
Furthermore, the same framework can be applied to investigate the fundamental mechanical properties of heavy mesons containing charm or bottom quarks by treating them as spinless matter fields.
Recent model-dependent calculations of the GFFs for quarkonium states ($\eta_c^{(\prime)}$, $\chi_{c0}$, $\eta_b$, as well as the hypothetical strangenium $\eta_s$)~\cite{Xu:2024hfx,Hu:2024edc,Sultan:2024hep}, which are ${\rm SU}(N)$-flavor singlet matter fields, further motivate a systematic effective field theory description of matter fields in curved spacetime.

We perform a systematic one-loop renormalization at $\mathcal{O}(p^3)$ by explicitly calculating the ultraviolet (UV) divergences of the generating functional.
Our calculation employs the background field method~\cite{Honerkamp:1971sh, Honerkamp:1972fd,Abbott:1981ke} together with the heat-kernel technique~\cite{Schwinger:1951nm, DeWitt:1975ys}.
The application of the heat-kernel technique within ChPT in flat spacetime was first introduced in Refs.~\cite{Gasser:1983yg,Gasser:1984gg}.
Subsequently, it has been extended to high-order calculations~\cite{Unterdorfer:2002zg,Bijnens:1999hw}, (virtual) photon--meson systems~\cite{Urech:1994hd,Knecht:1999ag,Schweizer:2002ft,Agadjanov:2013lra}, meson--baryon systems~\cite{Gasser:1987rb,Muller:1996vy,Meissner:1998rw}, and the Higgs effective field theory~\cite{Guo:2015isa,Buchalla:2017jlu, Alonso:2017tdy}.
The renormalization of ChPT at the leading one-loop order with a spinless matter field in the ${\rm SU}(N)$ fundamental representation in the flat spacetime was performed using the heat-kernel technique in Ref.~\cite{Du:2016ntw}, and a systematic subtraction of power counting breaking terms was performed in Ref.~\cite{Du:2016xbh} by calculating the corresponding part of the one-loop generating functional.
We emphasize that the renormalization of ChPT in curved spacetime for scalar matter fields not only establishes a self-consistent framework for the aforementioned applications, but also lays the essential technical foundation for extending this formalism to the nucleon sector.
The inclusion of nucleon fields introduces additional complexities, such as the fermionic nature of nucleons, which necessitates a super-matrix formulation~\cite{Neufeld:1998js}, and the presence of derivative couplings, which require a non-trivial extension of the standard heat-kernel formalism.
These challenges will be addressed in our forthcoming work~\cite{pion-nucleon}, for which the present study serves as a necessary precursor.

This paper is organized as follows.
In Sec.~\ref{Sec: building blocks}, we introduce the basic building blocks and specify our conventions.
In Sec.~\ref{Sec: chiral Lagrangians in curved spacetime}, we construct the complete chiral Lagrangians involving the Goldstone bosons and the spinless matter fields up to $\mathcal{O}(p^3)$, including the newly introduced curvature-induced operators $\mathcal{L}_{\phi{P}}^{(2,3|R)}$.
In Sec.~\ref{Sec: renormalization}, we perform a systematic one-loop renormalization of the generating functional and determine the $\beta$-function coefficients for all relevant LECs.
Section~\ref{Sec: summary} presents a brief summary.
Appendices~\ref{App: fluctuation expansion},~\ref{App: SU(N) Lie algebra relations}, and~\ref{App: field strength tensor} provide auxiliary details, including the second-order fluctuation expansion of the building blocks, the ${\rm SU}(N)$ Lie algebra relations, and the matrix elements of the field-strength tensor, respectively.

\section{Building Blocks} \label{Sec: building blocks}

The spontaneous breaking of the chiral group $G = {{\rm SU}(N)}_L \times {{\rm SU}(N)}_R$ to its vector subgroup $H = {{\rm SU}(N)}_V$ is nonlinearly realized by the Goldstone boson fields $\phi^a$ ($a = 1,\cdots,N^2-1$)~\cite{Coleman:1969sm, Callan:1969sn}:
\begin{align}
&   u_L(\phi) \xrightarrow[]{g} g_Lu_L(\phi){K(g, \phi)}^\dagger \: , \quad g = (g_L, g_R) \in G \: , \label{Eq: chiral transformation|u_L} \\
&   u_R(\phi) \xrightarrow[]{g} g_Ru_R(\phi){K(g, \phi)}^\dagger \: , \label{Eq: chiral transformation|u_R}
\end{align}
where $u_L$ and $u_R$ are elements of the coset space $G / H$, and the compensator field $K(g, \phi) \in H$ depends on both the chiral transformation $g$ and the Goldstone boson fields $\phi$.

The building blocks associated with the Goldstone boson fields involved in this work are
\begin{align}
   u_\mu &= i\left[{
         u_R^\dagger(\partial_\mu-ir_\mu)u_R
        -u_L^\dagger(\partial_\mu-il_\mu)u_L
    }\right] \: , \\
   \chi_\pm &= {u_R^\dagger}\chi{u_L} \pm {u_L^\dagger}\chi^\dagger{u_R} \: , \\
   f_{\pm}^{\mu\nu} &= {u_L^\dagger}{F_{L}^{\mu\nu}}{u_L} \pm {u_R^\dagger}{F_{R}^{\mu\nu}}{u_R} \: ,
\end{align}
where
\begin{align}
   r_\mu &= v_\mu + a_\mu \: , \quad
    l_\mu = v_\mu - a_\mu\:,\quad \chi = 2B(s + ip) \: , \\
   F_{R}^{\mu\nu} &= \partial^\mu{r^\nu} - \partial^\nu{r^\mu} - i\commute{r^\mu}{r^\nu} \: , \quad
    F_{L}^{\mu\nu} = \partial^\mu{l^\nu} - \partial^\nu{l^\mu} - i\commute{l^\mu}{l^\nu} \: .
\end{align}
Here, $v_\mu$, $a_\mu$, $s$, and $p$ denote the external vector, axial-vector, scalar, and pseudoscalar sources, respectively, and $B$ is a positive constant proportional to the quark condensate in the chiral limit.
All these building blocks $O = u_\mu, \chi_\pm, f_{\pm}^{\mu\nu}$ transform under the chiral group according to
\begin{equation}
O \xrightarrow[]{g} KOK^\dagger \: ,
\end{equation}
and their covariant derivatives are defined as
\begin{equation}
\nabla_\mu{O} = \nabla_\mu^\text{gr}{O} + \commute{\Gamma_\mu}{O} \: ,
\end{equation}
where the purely gravitational covariant derivative $\nabla_\mu^\text{gr}$ acting on a tensor is given by
\begin{equation}
    \nabla_\mu^\text{gr}{T_{\rho\sigma\cdots}^{\alpha\beta\cdots}}
=   \partial_\mu{T_{\rho\sigma\cdots}^{\alpha\beta\cdots}}
+   \Gamma_{\mu\nu}^{\alpha}{T_{\rho\sigma\cdots}^{\nu\beta\cdots}}
+   \cdots
-   \Gamma_{\mu\rho}^{\nu}{T_{\nu\sigma\cdots}^{\alpha\beta\cdots}}
-   \cdots \: .
\end{equation}
The Levi-Civita connection $\Gamma_{\alpha\beta}^{\lambda}$ and the chiral connection $\Gamma_\mu$ take the standard forms:
\begin{align}
   \Gamma_{\alpha\beta}^{\lambda}
&=   \frac{1}{2}g^{\lambda\sigma}
    \left({
         \partial_\alpha{g_{\beta\sigma}}
        +\partial_\beta{g_{\alpha\sigma}}
        -\partial_\sigma{g_{\alpha\beta}}
    }\right) , \\
   \Gamma_\mu
&=   \frac{1}{2}
    \left[{
         u_R^\dagger(\partial_\mu-ir_\mu)u_R
        +u_L^\dagger(\partial_\mu-il_\mu)u_L
    }\right]  .
\end{align}

The transformation properties of the spinless matter fields $P_i$ ($i = 1,\cdots,N$) under the chiral group are not uniquely fixed.
However, it is convenient to define them such that~\cite{Coleman:1969sm, Callan:1969sn, Georgi:1984zwz}
\begin{equation}
P \xrightarrow[]{g} PK^\dagger \: .
\end{equation}
The covariant derivative acting on the spinless matter fields is then defined as
\begin{equation}
\nabla_\mu{P} = \partial_\mu{P} + {P}\Gamma_\mu^\dagger \: .
\end{equation}

In addition to the chiral building blocks discussed above, we shall also make use of the Riemann tensor, the Ricci tensor, and the Ricci scalar. Our conventions for these curvature tensors are as follows:
\begin{equation}
    {R\indices{^{\rho}_{\sigma\mu\nu}}}
=   \partial_\mu{\Gamma\indices{^{\rho}_{\nu\sigma}}}
-   \partial_\nu{\Gamma\indices{^{\rho}_{\mu\sigma}}}
+   {\Gamma\indices{^{\rho}_{\mu\lambda}}}{\Gamma\indices{^{\lambda}_{\nu\sigma}}}
-   {\Gamma\indices{^{\rho}_{\nu\lambda}}}{\Gamma\indices{^{\lambda}_{\mu\sigma}}} \: , \quad
    {R\indices{_{\mu\nu}}}
=   {R\indices{^{\alpha}_{\mu\alpha\nu}}} \: , \quad
    {R}
=   g^{\mu\nu}{R\indices{_{\mu\nu}}} \: .
\end{equation}

\section{Chiral Lagrangians in Curved Spacetime} \label{Sec: chiral Lagrangians in curved spacetime}

We restrict ourselves to the single spinless matter sector.
The relevant generating functional is defined as
\begin{equation}
    e^{iZ[j, J, J^\dagger]}
=   \int{[d\phi][dP][dP^\dagger]}\:{
        \exp\left\{{
            i\int{d^4x}\:{
                \sqrt{-g}
                \left[{
                     \mathcal{L}_{\phi}(j)
                    +\mathcal{L}_{\phi{P}}
                    +\left({
                        {P}{J^\dagger} + {J}{P^\dagger}
                    }\right)
                }\right]
            }
        }\right\}
    } \: ,
\end{equation}
where $J$ and $J^\dagger$ denote external sources coupled to the spinless matter fields, and $j = \left\{{v_\mu, a_\mu, s, p, g_{\mu\nu}}\right\}$ collectively represents all remaining external sources.
The chiral Lagrangians in curved spacetime can be organized according to their chiral dimensions as
\begin{equation}
    \mathcal{L}_{\phi}
=   \sum_{n=1}^{\infty}\left(
         \mathcal{L}_{\phi}^{(2n|M)}
        +\mathcal{L}_{\phi}^{(2n|R)}
    \right) , \quad
    \mathcal{L}_{\phi{P}}
=   \sum_{n=1}^{\infty} \left(
         \mathcal{L}_{\phi{P}}^{(n|M)}
        +\mathcal{L}_{\phi{P}}^{(n|R)}
    \right) ,
\end{equation}
where $n$ denotes the chiral dimension. The superscripts $M$ and $R$ indicate, respectively, the terms obtained by extending the flat-spacetime Lagrangians to curved spacetime through minimal coupling, and the additional curvature-induced contributions that arise specifically in curved spacetime.

For the Goldstone boson sector, the flat-spacetime contributions up to $\mathcal{O}(p^4)$ take the form~\cite{Gasser:1983yg, Gasser:1984gg}
\begin{align}
    \mathcal{L}_{\phi}^{(2|M)}
=&\,   \frac{F^2}{4}\flavor{{u_\mu}{u^\mu}}
    +\frac{F^2}{4}\flavor{\chi_+} \: , \\
    \mathcal{L}_{\phi}^{(4|M)}
=&\,   L_{0}\flavor{{u_\mu}{u_\nu}{u^\mu}{u^\nu}}
    +L_{1}\flavor{{u_\mu}{u^\mu}}\flavor{{u_\nu}{u^\nu}}
    +L_{2}\flavor{{u_\mu}{u_\nu}}\flavor{{u^\mu}{u^\nu}}
    +L_{3}\flavor{{u_\mu}{u^\mu}{u_\nu}{u^\nu}} \nonumber\\
 &\,  +L_{4}\flavor{{u_\mu}{u^\mu}}\flavor{\chi_+}
    +L_{5}\flavor{{u_\mu}{u^\mu}{\chi_+}}
    +L_{6}{\flavor{\chi_+}}^{2}
    +L_{7}{\flavor{\chi_-}}^{2}
    +\frac{1}{2}L_{8}\flavor{{\chi_+^2} + {\chi_-^2}} \nonumber\\
 &\,  -iL_{9}\flavor{{u_\mu}{u_\nu}{f_{+}^{\mu\nu}}}
    +\frac{1}{4}L_{10}\flavor{{f_+^2} - {f_-^2}}
    +H_{1}\flavor{{F_L^2} + {F_R^2}}
    +H_{2}\flavor{{\chi}{\chi^\dagger}} \: ,
\end{align}
where $F$ is the pion decay constant in the chiral limit.
The curvature-induced terms were first systematically constructed in Ref.~\cite{Donoghue:1991qv}:
\begin{align}
    \mathcal{L}_{\phi}^{(2|R)}
=&\,  H_{0}R \: , \\
    \mathcal{L}_{\phi}^{(4|R)}
=&\,   L_{11}R\flavor{{u_\mu}{u^\mu}}
    +L_{12}R^{\mu\nu}\flavor{{u_\mu}{u_\nu}}
    +L_{13}R\flavor{\chi_+} \nonumber\\
 &\,  +H_{3}R^2
    +H_{4}R^{\mu\nu}R_{\mu\nu}
    +H_{5}R^{\mu\nu\alpha\beta}R_{\mu\nu\alpha\beta} \: .
\end{align}
Note that the Riemann tensor $R^{\mu\nu\alpha\beta}$, the Ricci tensor $R^{\mu\nu}$, and the Ricci scalar $R$ are constructed from two derivatives of the metric tensor and thus carry chiral dimension $p^2$.

The first class of contributions to the single spinless-matter sector up to $\mathcal{O}(p^3)$ is given by~\cite{Roessl:1999iu, Guo:2008gp, Yao:2015qia}
\begin{align}
    \mathcal{L}_{\phi{P}}^{(1|M)}
=&\,   \nabla_\mu{P}\nabla^\mu{P^\dagger}
    -m^2{P}{P^\dagger} \: , \\
    \mathcal{L}_{\phi{P}}^{(2|M)}
=&\,   P\left[{
        -h_{0}\flavor{\chi_+}
        -h_{1}\chi_+
        +h_{2}\flavor{{u_\mu}{u^\mu}}
        -h_{3}{u_\mu}{u^\mu}
    }\right]P^\dagger \nonumber\\
 &\,  +\nabla_\mu{P}\left[{
         h_{4}\flavor{{u^\mu}{u^\nu}}
        -h_{5}\anticommute{u^\mu}{u^\nu}
    }\right]\nabla_\nu{P^\dagger} \: , \\
    \mathcal{L}_{\phi{P}}^{(3|M)}
=&\,  \Big[{
         ig_{1}{P}\commute{\chi_-}{u_\mu}\nabla^\mu{P^\dagger}
        +g_{2}{P}\commute{u^\mu}{\nabla_\mu{u_\nu} + \nabla_\nu{u_\mu}}\nabla^\nu{P^\dagger}
    }\Big. \nonumber\\
 &\,  \Big.{
        +g_{3}{P}\commute{u_\mu}{\nabla_\alpha{u_\beta}}\anticommute{\nabla^\mu}{\anticommute{\nabla^\alpha}{\nabla^\beta}}{P^\dagger}
        +g_{4}{P}\left({\nabla_\mu{\chi_+}}\right)\nabla^\mu{P^\dagger}
    }\Big. \nonumber\\
 &\,  \Big.{
        +g_{5}{P}\flavor{\nabla_\mu{\chi_+}}\nabla^\mu{P^\dagger}
        +{\rm h.c.}
    }\Big]
        +i\gamma_{1}\nabla^\mu P f^{+}_{\mu\nu}\nabla^\nu{P^\dagger}+\gamma_{2}{P}\commute{u^\mu}{f_{\mu\nu}^{-}}\nabla^\nu{P^\dagger} \: ,
\end{align}
where $m$ denotes the mass of the spinless matter field in the chiral limit.
In what follows, we construct the second class of contributions,\footnote{For pedagogical treatments of the construction of the most general, linearly independent chiral Lagrangians, we refer the reader to Refs.~\cite{Fearing:1994ga,Bijnens:1999sh}.} namely the curvature-induced terms.
At $\mathcal{O}(p^2)$, the most general structure is obtained by taking all possible full contractions of $R^{\mu\nu\rho\sigma}$, $R^{\mu\nu}$, or $R$ with $\mathcal{O}(p^0)$ operators of the form ${P}\nabla_\alpha\cdots\nabla_\beta{P^\dagger}$.
Using the identity
\begin{align}\label{Eq:commutator}
    \commute{\nabla_\mu}{\nabla_\nu}\nabla_{\alpha_1}\cdots\nabla_{\alpha_r}{P^\dagger}
=&   \left({
         \frac{1}{4}\commute{u_\mu}{u_\nu}
        -\frac{i}{2}f_{\mu\nu}^{+}
    }\right)
    \nabla_{\alpha_1}\cdots\nabla_{\alpha_r}{P^\dagger} \notag \\
 &  -\sum_{i=1}^{r}\:{
        {R\indices{^{\lambda}_{\alpha_i\mu\nu}}}
        \nabla_{\alpha_1}\cdots\nabla_{\lambda}\cdots\nabla_{\alpha_r}{P^\dagger}
    } \: ,
\end{align}
it follows that the antisymmetric part of the covariant derivatives contributes only at higher orders; see also Ref.~\cite{Guo:2008gp}.
Therefore, only the totally symmetrized covariant derivatives acting on the spinless matter fields need to be retained:
\begin{equation}
    \nabla_{\{\alpha_1}\cdots\nabla_{\alpha_r\}}{P^\dagger}
\equiv   \frac{1}{r!}
    \sum_{\mathcal{P}}\:{
        \nabla_{\alpha_{\mathcal{P}_1}}\cdots\nabla_{\alpha_{\mathcal{P}_r}}{P^\dagger}
    } \: ,
\end{equation}
where the sum runs over all permutations $\mathcal{P}$ of the indices $\{1,\ldots,r\}$.
Furthermore, taking into account the antisymmetry of the Riemann tensor under the exchange of indices within each pair, the minimal operator basis at this order reads
\begin{equation}\label{Eq: L_R2}
    \mathcal{L}_{\phi{P}}^{(2|R)}
=    d_{1}R^{\mu\nu}{P}\nabla_{\{\mu}\nabla_{\nu\}}{P^\dagger}
    +d_{2}R{P}{P^\dagger} \: .
\end{equation}
At $\mathcal{O}(p^3)$, the most general structures involve all possible full contractions of $\nabla_\lambda{R^{\mu\nu}}$ and $\nabla_\lambda{R}$ with ${P}\nabla_{\{\alpha}\cdots\nabla_{\beta\}}{P^\dagger}$.\footnote{Terms containing factor of $\nabla_\rho{R\indices{^\rho_\mu}}$ can be reduced according to the second Bianchi identity, $\nabla_\rho{R\indices{^\rho_\mu}} = \frac{1}{2}\nabla_\mu{R}$.}
For the same reason as above, terms containing $\nabla_\lambda{R^{\mu\nu\rho\sigma}}$ do not need to be included.\footnote{Note that $R_{\rho \sigma \mu \nu}=-R_{\sigma \rho \mu \nu}$ and $R_{\rho \sigma \mu \nu}=-R_{\rho \sigma \nu \mu}$.}
In addition, the axial-vector vielbein $u_\mu$ is excluded by $P$-parity considerations.
The minimal operator basis at this order therefore takes the form
\begin{equation}\label{Eq: L_R3}
    \mathcal{L}_{\phi{P}}^{(3|R)}
=    e_{1}\left({\nabla^\lambda{R^{\mu\nu}}}\right){P}\nabla_{\{\lambda}\nabla_{\mu}\nabla_{\nu\}}{P^\dagger}
    +e_{2}\left({\nabla^\mu{R}}\right){P}\nabla_\mu{P^\dagger} \: .
\end{equation}

\section{Renormalization} \label{Sec: renormalization}

The generating functional up to $\mathcal{O}(p^3)$ consists of two parts,
\begin{equation} \label{Eq: 3rd order Z}
Z_3 = Z_3^\text{tree} + Z_3^\text{one-loop} \: .
\end{equation}
The tree-level contribution takes the form
\begin{equation}
    Z_3^\text{tree}
=   \int{d^4x}\:{
        \sqrt{-g}
        \left({
             \mathcal{L}_{\phi}^{(2|M)}
            +\mathcal{L}_{\phi{P}}^{(1|M)}
            +\mathcal{L}_{\phi{P}}^{(2|M,R)}
            +\mathcal{L}_{\phi{P}}^{(3|M,R)}
        }\right)
    } \: .
\end{equation}
The one-loop contribution is evaluated using the background field method.
In this approach, the fields appearing in the Lagrangians are decomposed into classical (background) and quantum (fluctuation) components,
\begin{equation}
\phi = \bar{\phi} + \varphi \: , \quad
P = \bar{P} + h \: ,
\end{equation}
where the classical configurations are determined by the external sources through the lowest-order equations of motion:
\begin{equation}
\nabla_\mu{u^\mu} = \frac{i}{2}\left({\chi_- - \frac{1}{N}\flavor{\chi_-}}\right) \: , \quad
\nabla_\mu\nabla^\mu{P} + m^2{P} = 0 \: .
\end{equation}
Working in the standard chiral gauge, defined by $u_R(\bar{\phi}) = u_L^\dagger(\bar{\phi}) \equiv \bar{u}$, it is convenient to parameterize the Goldstone-boson fluctuations as
\begin{equation}
u_R = \bar{u}\,e^{i\frac{\eta}{2F}} \: , \quad
u_L = \bar{u}^{\dagger}\,e^{-i\frac{\eta}{2F}} \: ,
\end{equation}
where $\eta = \eta^a\lambda^a$ with $a = 1,\ldots,N^2-1$.
In the following, we omit the bars on classical fields for notational simplicity.
Collecting the fluctuation variables into
\begin{equation}
\xi_A = \left({\eta^a, \sqrt{2}h_i}\right) \: ,
\end{equation}
with $i =1, \ldots, N$, 
the one-loop functional in Eq.~\eqref{Eq: 3rd order Z} can be expressed as a Gaussian integral,
\begin{equation} \label{Eq: one-loop Z|Gaussian integral}
    e^{iZ_3^\text{one-loop}}
=   \int{[d\xi]}\:{
        \exp\left\{{
            -\frac{i}{2}
            \int{d^4x}\:{
                \sqrt{-g}\,
                \xi_A
                {\left({
                     \frac{1}{\sqrt{-g}}\mathbb{D}_\mu\sqrt{-g}g^{\mu\nu}\mathbb{D}_\nu
                    +\mathbb{Q}
                }\right)}^{AB}
                \xi_B^\dagger
            }
        }\right\}
    } \: .
\end{equation}
Here, the covariant derivative $\mathbb{D}_{\mu}^{AB}$ is defined as\footnote{The formulas for the second-order fluctuation expansion of the building blocks and the $\mathrm{SU}(N)$ Lie algebra relations used here are given in Appendices~\ref{App: fluctuation expansion} and~\ref{App: SU(N) Lie algebra relations}, respectively.}
\begin{equation}
    \mathbb{D}_{\mu}^{AB}
=   \delta^{AB}\partial_\mu
+   \mathbb{X}_{\mu}^{AB} \: ,
\end{equation}
where the derivative-free connection matrix $\mathbb{X}_{\mu}^{AB}$ takes the form
\begin{equation}
    \mathbb{X}_{\mu}^{AB}
=   \begin{pmatrix}
        -\frac{1}{2}\flavor{\Gamma_\mu\commute{\lambda^a}{\lambda^b}}
        -\frac{1}{8F^2}\big({
             \nabla_\mu{P}\commute{\lambda^a}{\lambda^b}{P^\dagger}
            +{\rm h.c.}
        }\big)
        &
        \frac{1}{4\sqrt{2}F}{\big({P\commute{u_\mu}{\lambda^a}}\big)}_{j}
        \\
        \frac{1}{4\sqrt{2}F}{\big({\commute{u_\mu}{\lambda^b}P^\dagger}\big)}_{i}
        &
        {\big({\Gamma_\mu}\big)}_{ij}
    \end{pmatrix} \: .
\end{equation}
The matrix elements of $\mathbb{Q}^{AB}$ are given by~\cite{Du:2016ntw}
\begin{align}
&   \begin{aligned}
    \mathbb{Q}^{ab}
=&   \frac{1}{8}\flavor{u_\mu\commute{\commute{u^\mu}{\lambda^a}}{\lambda^b}}
    +\frac{1}{16}\flavor{\anticommute{\anticommute{\chi_+}{\lambda^a}}{\lambda^b}}
    +\frac{3}{32F^2}{P}\commute{u_\mu}{\lambda^a}\commute{u^\mu}{\lambda^b}{P^\dagger} \\
 &  -\frac{1}{64F^4}\big({
         \nabla_\mu{P}\commute{\lambda^a}{\lambda^c}{P^\dagger}
        +{\rm h.c.}
    }\big)\big({
         \nabla^\mu{P}\commute{\lambda^c}{\lambda^b}{P^\dagger}
        +{\rm h.c.}
    }\big) \: ,
\end{aligned} \notag \\
&   \begin{aligned}
    \mathbb{Q}^{aj}
=&  -\frac{1}{4\sqrt{2}F}{\big({{P}\commute{\nabla_\mu{u^\mu}}{\lambda^a}}\big)}^{j}
    -\frac{3}{4\sqrt{2}F}{\big({\nabla_\mu{P}\commute{u^\mu}{\lambda^a}}\big)}^{j} \\
 &  +\frac{1}{32\sqrt{2}F^3}\big({
         \nabla_\mu{P}\commute{\lambda^a}{\lambda^c}{P^\dagger}
        +{\rm h.c.}
    }\big){\big({
        {P}\commute{u^\mu}{\lambda^c}
    }\big)}^{j} \: ,
\end{aligned} \notag \\
&   \begin{aligned}
    \mathbb{Q}^{ib}
=&   \frac{1}{4\sqrt{2}F}{\big({\commute{\nabla_\mu{u^\mu}}{\lambda^b}{P^\dagger}}\big)}^{i}
    +\frac{3}{4\sqrt{2}F}{\big({\commute{u^\mu}{\lambda^b}\nabla_\mu{P^\dagger}}\big)}^{i} \\
 &  +\frac{1}{32\sqrt{2}F^3}{\big({
        \commute{u_\mu}{\lambda^c}{P^\dagger}
    }\big)}^{i}\big({
         \nabla^\mu{P}\commute{\lambda^c}{\lambda^b}{P^\dagger}
        +{\rm h.c.}
    }\big) \: ,
\end{aligned} \notag \\
&   \begin{aligned}
    \mathbb{Q}^{ij}
=&   m^2\delta^{ij}
    -\frac{1}{32F^2}{\big({
        \commute{u_\mu}{\lambda^c}{P^\dagger}
    }\big)}^{i}{\big({
        {P}\commute{u^\mu}{\lambda^c}
    }\big)}^{j} \: .
\end{aligned}
\end{align}

Evaluating the Gaussian integral in Eq.~\eqref{Eq: one-loop Z|Gaussian integral} yields
\begin{equation}
    Z_3^\text{one-loop}
=   \frac{i}{2}
    {\rm Tr}\:{
        \ln{\left[{
            \int{d^4x}\:{
                \sqrt{-g}
                |{x}\rangle
                {\left({
                     \frac{1}{\sqrt{-g}}\mathbb{D}_\mu\sqrt{-g}g^{\mu\nu}\mathbb{D}_\nu
                    +\mathbb{Q}
                }\right)}
                \langle{x}|
            }
        }\right]}
    } \: .
\end{equation}
Here we adopt the abstract Hilbert-space notation~\cite{Lee:1984ud, Parker:1978gh}, where the basis vectors $|{x}\rangle$ in curved spacetime satisfy
\begin{align}
&   \langle{x}|{x^\prime}\rangle = \frac{1}{\sqrt{-g}}\delta(x - x^\prime)
&&  \text{(orthogonality)} \: , \\
&   \int{d^4x}\:{\sqrt{-g}|{x}\rangle\langle{x}|} = \mathcal{I}
&&  \text{(completeness)} \: .
\end{align}
Furthermore, the full trace of an operator $\mathcal{O}$ is defined as
\begin{equation}
    {\rm Tr}\:{\mathcal{O}}
=   \int{d^4x}\:{
        \sqrt{-g}\,
        {\rm tr}\:{
            \langle{x}|{\mathcal{O}}|{x}\rangle
        }
    } \: ,
\end{equation}
where ``${\rm tr}$'' denotes the trace over the $(N^2 - 1 + N)$-dimensional internal space spanned by the basis ${\xi_A}$.
The UV divergences of $Z_3^\text{one-loop}$ can be extracted using the heat-kernel technique.
One finds~\cite{Donoghue:1991qv}
\begin{equation} \label{Eq: one-loop Z|divergence}
    {\rm div}\:{Z_3^\text{one-loop}}
=   -\frac{\mu^{d-4}}{{(4\pi)}^{d/2}}\frac{1}{d-4}
    \int{d^dx}\:{
        \sqrt{-g}
        \left\{{
             {\rm tr}\:\left[{
                 \frac{1}{12}\mathbb{F}_{\mu\nu}\mathbb{F}^{\mu\nu}
                +\frac{1}{2}\mathbb{Q}^2
                +\frac{1}{6}R\mathbb{Q}
            }\right]
            +\cdots
        }\right\}
    } \: ,
\end{equation}
where, for the purpose of this work, we consider only the single spinless-matter contribution on the right-hand side of Eq.~\eqref{Eq: one-loop Z|divergence}.
Dimensional regularization is employed throughout, with $\mu$ being the renormalization scale.
The matrix elements of the field-strength tensor $\mathbb{F}_{\mu\nu}^{AB} = {\commute{\mathbb{D}_\mu}{\mathbb{D}_\nu}}^{AB}$ are listed in Appendix~A of Ref.~\cite{Du:2016ntw} as well as in Appendix~\ref{App: field strength tensor} here.

In order to obtain a finite generating functional up to $\mathcal{O}(p^3)$, the UV divergences of $Z_3^\text{one-loop}$ must be removed by renormalizing the LECs appearing in $Z_3^\text{tree}$:
\begin{equation}
c_i = c_i^r(\mu) + \delta{c_i}\Lambda
\quad \text{with} \quad
\Lambda = \frac{\mu^{d-4}}{{(4\pi)}^{d/2}}\frac{1}{d-4} \: ,
\end{equation}
where $c_i^r(\mu)$ denotes the finite, scale- and scheme-dependent renormalized LEC, and $\delta{c_i}$ is the corresponding $\beta$-function coefficient.
The terms proportional to $\mathbb{F}_{\mu\nu}\mathbb{F}^{\mu\nu}$ and $\mathbb{Q}^2$ in Eq.~\eqref{Eq: one-loop Z|divergence} contribute to the renormalization of $\mathcal{L}_{\phi{P}}^{(2,3|M)}$, as previously derived in Ref.~\cite{Du:2016ntw}:
\begin{align}
   \delta{h_{0,1}} &= 0 \: , \quad
    \delta{h_{2}} = \frac{m^2}{24} \: , \quad
    \delta{h_{3}} = -\frac{m^2}{24}N \: , \quad
    \delta{h_{4}} = \frac{7}{12} \: , \quad
    \delta{h_{5}} = -\frac{7}{24}N \: , \notag \\
   \delta{g_{1,3,4,5}} &= 0 \: , \quad
    \delta{g_{2}} = -\frac{3}{64}N \: , \quad
    \delta{\gamma_{1}} = \frac{1}{6}N \: , \quad
    \delta{\gamma_{2}} = \frac{11}{96}N \: .
\end{align}
The curvature-induced term
\begin{equation}
    \frac{1}{6}R\,{\rm tr}\:{\mathbb{Q}}
=   -\frac{1}{48F^2}{R}{P}\flavor{{u_\mu}{u^\mu}}{P^\dagger}
    -\frac{1}{48F^2}\frac{N^2-2}{N}{R}{P}{u_\mu}{u^\mu}{P^\dagger}
    +\cdots \: ,
\end{equation}
which explicitly contains the Ricci scalar, is of $\mathcal{O}(p^4)$.
Consequently, the new LECs within $\mathcal{L}_{\phi{P}}^{(2,3|R)}$ in Eqs.~\eqref{Eq: L_R2} and~\eqref{Eq: L_R3} are free of UV divergent contributions, i.e.,
\begin{equation}\label{Eq: d0-4}
\delta{d_{1,2}} = \delta{e_{1,2}} = 0 \: .
\end{equation}
These $\beta$-function coefficients determine the scale dependence of the corresponding renormalized LECs through the renormalization group equations
\begin{equation}
\mu\frac{\partial{c_i^r(\mu)}}{\partial{\mu}} = -\frac{\delta{c_i}}{16\pi^2} \: .
\end{equation}

\section{Summary} \label{Sec: summary}

In this work, we have constructed the chiral Lagrangians in curved spacetime for spinless matter fields transforming in the fundamental representation of ${\rm SU}(N)$ up to the leading one-loop order in the chiral expansion, $\mathcal{O}(p^3)$, with particular emphasis on the curvature-induced terms.
Using the background field method and the heat-kernel technique, we have derived the ultraviolet divergences of the generating functional for correlation functions in the single spinless-matter sector to the same chiral order.
Our central result, as shown in Eq.~\eqref{Eq: d0-4}, is that the four curvature-induced low-energy constants $d_{1,2}$ and $e_{1,2}$ introduced in this work are ultraviolet finite, analogous to $c_9$ in the $\pi N$ system~\cite{Alharazin:2020yjv}.

This work establishes the essential theoretical foundation for model-independent investigations of the graviproduction of kaons and the gravitational form factors of heavy mesons.
Furthermore, the techniques and results presented here serve as a necessary precursor for our ongoing efforts to address the more involved problem of renormalizing chiral perturbation theory with nucleon fields in curved spacetime~\cite{pion-nucleon}.

\section*{Acknowledgments}

This work is supported in part by National Natural Science Foundation of China under Grants No.~12125507 and No.~12447101; by the Chinese Academy of Sciences under Grant No.~YSBR-101; and by the Postdoctoral Fellowship Program of China Postdoctoral Science Foundation under Grants No.~GZC20232773 and No.~2023M74360. 

\medskip

\appendix

\section{Second-Order Fluctuation Expansion of Building Blocks} \label{App: fluctuation expansion}

To second order in the fluctuation field variable $\eta$, the chiral connection $\Gamma_\mu$, the axial-vector vielbein $u_\mu$, and $\chi_\pm$ are given by
\begin{equation}
\begin{aligned}
    \Gamma_\mu
&=  \bar{\Gamma}_\mu
 +  \frac{1}{4F}\commute{\bar{u}_\mu}{\eta}
 -  \frac{1}{8F^2}\commute{\bar{\nabla}_\mu\eta}{\eta}
 +  \mathcal{O}(\eta^3) \: , \\
    u_\mu
&=  \bar{u}_\mu
 -  \frac{1}{F}\bar{\nabla}_\mu\eta
 -  \frac{1}{8F^2}\commute{\commute{\bar{u}_\mu}{\eta}}{\eta}
 +  \mathcal{O}(\eta^3) \: , \\
    \chi_\pm
&=  \bar{\chi}_\pm
 -  \frac{i}{2F}\anticommute{\bar{\chi}_\mp}{\eta}
 -  \frac{1}{8F^2}\anticommute{\anticommute{\bar{\chi}_\pm}{\eta}}{\eta}
 +  \mathcal{O}(\eta^3) \: ,
\end{aligned}
\end{equation}
where the chiral covariant derivative of the fluctuation field is defined as $\bar{\nabla}_\mu\eta = \partial_\mu\eta + \commute{\bar{\Gamma}_\mu}{\eta}$.

\section{${\rm SU}(N)$ Lie Algebra Relations} \label{App: SU(N) Lie algebra relations}

Regarding the generators $\lambda^a$ ($a = 1,\cdots,N^2-1$) of the Lie group ${\rm SU}(N)$ in the fundamental representation, we adopt the normalization convention
\begin{equation}
\flavor{\lambda^a\lambda^b} = 2\delta^{ab} \: .
\end{equation}
These generators satisfy the commutation and anticommutation relations of the Lie algebra:
\begin{equation}
\commute{\lambda^a}{\lambda^b} = 2if^{abc}\lambda^c \: , \quad
\anticommute{\lambda^a}{\lambda^b} = \frac{4}{N}\delta^{ab}I + 2d^{abc}\lambda^c \: ,
\end{equation}
where $f^{abc}$ are the completely antisymmetric structure constants and $d^{abc}$ are the completely symmetric tensors.
These satisfy the orthogonality relations
\begin{equation}
f^{acd}f^{bcd} = 2N\delta^{ab} \: , \quad
d^{acd}d^{bcd} = \frac{2(N^2-4)}{N}\delta^{ab} \: , \quad
f^{acd}d^{bcd} = 0 \: .
\end{equation}
Moreover, the Fierz completeness relation states that an arbitrary $N \times N$ matrix $M$ can be expanded in terms of the basis $\{I,\lambda^a\}$:
\begin{equation}
    M
=   \frac{1}{N}\flavor{M}{I}
+   \frac{1}{2}\flavor{M\lambda^a}\lambda^a \: ,
\end{equation}
or equivalently, that the following identity holds when acting on the tensor product of two fundamental representation spaces:
\begin{equation}
    {\big({\lambda^a}\big)}_{ij}{\big({\lambda^a}\big)}_{kl}
=   2\left({
         \delta_{il}\delta_{jk}
        -\frac{1}{N}\delta_{ij}\delta_{kl}
    }\right) \: .
\end{equation}

\section{Matrix Elements of Field Strength Tensor} \label{App: field strength tensor}

The field strength tensor in Eq.~\eqref{Eq: one-loop Z|divergence} is
\begin{equation}
    \mathbb{F}_{\mu\nu}^{AB}
=   \begin{pmatrix}
        -\frac{1}{2}\flavor{
            \left({
                \Gamma_{\mu\nu} + A_{\mu\nu}
            }\right)
            \commute{\lambda^a}{\lambda^b}
        }
        +\Sigma_{\mu\nu}^{ab}
        &
        \Sigma_{\mu\nu}^{aj}
        \\
        \Sigma_{\mu\nu}^{ib}
        &
        \Gamma_{\mu\nu}^{ij} + \Sigma_{\mu\nu}^{ij}
    \end{pmatrix} \: ,
\end{equation}
where the various components are defined as follows:
\begin{align}
    \Gamma_{\mu\nu}
& =  \partial_\mu{\Gamma_\nu}
    -\partial_\nu{\Gamma_\mu}
    +\commute{\Gamma_\mu}{\Gamma_\nu} \: ,
\notag \\
    A_{\mu\nu}
&= \frac{1}{4F^2}\big({
         2\nabla_\mu{P^\dagger}\nabla_\nu{P}
        -2\nabla_\nu{P^\dagger}\nabla_\mu{P}
        +{P^\dagger}\commute{\nabla_\mu}{\nabla_\nu}{P}
        -\commute{\nabla_\mu}{\nabla_\nu}{P^\dagger}{P}
    }\big) \notag \\
 & \quad +\frac{1}{16F^4}\commute{
        {P^\dagger}\nabla_\mu{P} - \nabla_\mu{P^\dagger}{P}
    }{
        {P^\dagger}\nabla_\nu{P} - \nabla_\nu{P^\dagger}{P}
    } \: ,
 \notag \\
    \Sigma_{\mu\nu}^{ab}
& = \frac{1}{32F^2}\Big({
         {P}\commute{u_\mu}{\lambda^a}\commute{u_\nu}{\lambda^b}{P^\dagger}
        -{P}\commute{u_\nu}{\lambda^a}\commute{u_\mu}{\lambda^b}{P^\dagger}
    }\Big) \: , \notag \\
    \Sigma_{\mu\nu}^{aj}
&= \frac{1}{4\sqrt{2}F}{\big({
         \nabla_\mu{P}\commute{u_\nu}{\lambda^a}
        -\nabla_\nu{P}\commute{u_\mu}{\lambda^a}
        +{P}\commute{\nabla_\mu{u_\nu} - \nabla_\nu{u_\mu}}{\lambda^a}
    }\big)}_{j} \notag \\
 & \quad -\frac{1}{32\sqrt{2}F^3}\left[{
         \big({
             \nabla_\mu{P}\commute{\lambda^a}{\lambda^c}{P^\dagger}
            -{P}\commute{\lambda^a}{\lambda^c}\nabla_\mu{P^\dagger}
        }\big){\big({
            {P}\commute{u_\nu}{\lambda^c}
        }\big)}_{j}
        -\big({\mu \leftrightarrow \nu}\big)
    }\right] \: , \notag \\
    \Sigma_{\mu\nu}^{ib}
& = \frac{1}{4\sqrt{2}F}{\Big({
         \commute{u_\nu}{\lambda^b}\nabla_\mu{P^\dagger}
        -\commute{u_\mu}{\lambda^b}\nabla_\nu{P^\dagger}
        +\commute{\nabla_\mu{u_\nu} - \nabla_\nu{u_\mu}}{\lambda^b}{P^\dagger}
    }\Big)}_{i} \notag \\
 & \quad -\frac{1}{32\sqrt{2}F^3}\left[{
         {\big({
            \commute{u_\mu}{\lambda^c}{P}^\dagger
        }\big)}_{i}\Big({
             \nabla_\nu{P}\commute{\lambda^c}{\lambda^b}{P^\dagger}
            -{P}\commute{\lambda^c}{\lambda^b}\nabla_\nu{P^\dagger}
        }\Big)
        -\big({\mu \leftrightarrow \nu}\big)
    }\right] \: , \notag \\
    \Sigma_{\mu\nu}^{ij}
&= \frac{1}{32\sqrt{2}F^2}\left[{
         {\big({
            \commute{u_\mu}{\lambda^c}{P}^\dagger
        }\big)}_{i}{\big({
            {P}\commute{u_\nu}{\lambda^c}
        }\big)}_{j}
        -{\big({
            \commute{u_\nu}{\lambda^c}{P}^\dagger
        }\big)}_{i}{\big({
            {P}\commute{u_\mu}{\lambda^c}
        }\big)}_{j}
    }\right] \: .
\end{align}

\bibliographystyle{jhep}
\bibliography{refs}

\end{document}